\documentclass[english,preprint]{revtex4-1}
\usepackage[T1]{fontenc}
\usepackage[latin9]{inputenc}
\usepackage{amsmath}
\usepackage{graphicx}
\usepackage{esint}
\usepackage{color}
\makeatletter

\usepackage{babel}

\makeatother

\usepackage{babel}
\begin{document}
\pagecolor{white}
\title{Passage through a sub-diffusing geometrical bottleneck}
\author{K L Sebastian}
\email{email: kls@iitpkd.ac.in}
\address{Indian Institute of Technology, Palakkad\\ Ahalia Integrated Campus, Kozhippara P.O.\\ Palakkad 678557,India\\ \bf (To appear in The Journal of Chemical Physics)}
\begin{abstract} 
The usual Kramers theory of reaction rates in  condensed media predict the rate to have an inverse dependence on the viscosity of the medium, $\eta$.  However, experiments on ligand binding to proteins, performed long ago,  showed the rate to have $\eta^{-\nu}$ dependence, with $\nu $ in the range $0.4-0.8$.   Zwanzig {\it (Journal of Chemical Physics 97, 3587 (1992))} suggested a model, in which the ligand has to pass through a fluctuating opening to reach the binding site.  This fluctuating gate model predicted the rate to be proportional to $\eta^{-1/2}$.    More recently, experiments performed by Xie et. al. ({\it  Physical Review Letters 93, 1 (2004)}) showed that the distance between two groups in a protein undergoes not normal diffusion, but subdiffusion.  Hence in this paper, we suggest and solve a generalisation of the Zwanzig model, viz., passage through an opening, whose size undergoes sub-diffusion.   Our solution shows that the rate is proportional to $\eta^{-\nu}$ with $\nu$ in the range $0.5-1$, and hence the sub-diffusion model can explain the experimental observations. 

 \end{abstract}
\maketitle

\section{Introduction}

Simple, exactly solvable models of chemical reaction dynamics are
very useful, as they give very valuable insights into the process.
Among the very few such models is the one due to Zwanzig \citep{Zwanzig1992},
for the passage of a ligand molecule through a fluctuating bottleneck.
Many authors have suggested similar models for the removal of a steric
constraint by fluctuations, for molecular rotation in liquids and
glasses \citep{Glarum1960,Hunt1966,Bordewijk1975,Frobose1986,Klafter1985}.
The model of Zwanzig is for the passage of a ligand to a binding site
that is buried deep inside a cavity within a protein. It assumes that
the rate of binding is proportional to the area of the opening, which
undergoes time dependent fluctuations. In general, the predictions of the model
are in agreement with the experiments. The concentration
of the ligand initially decays non-exponentialy, but changes over
to exponential at long times. Taking the time scale of decay of the
fluctuations of the opening as proportional to the viscosity $\eta$
of the medium, the model predicted that the rate constant for long
term decay is $\propto$ $\eta^{-1/2}$. In comparison, the Kramers
theory of activated processes leads to a rate proportional to $\eta^{-1}$.
The experiments of Beece \emph{et al.} \citep{Beece-1980} on the viscosity
dependence of the rate, found an inverse fractional dependence of
the form $\eta^{-\nu}$, in agreement with the Zwanzig theory. However,
the value of $\nu$ was in the range $0.4-0.8$, prompting the study
by Wang and Wolynes \citep{Wolynes1993}, of its extension to a non-Markovian
model, in which the relaxation of the opening was taken to be a stretched exponential.
They used the path integral technique to obtain the exact solution in
the long time limit. Their model leads to $\nu$ values that are strictly
less than $1/2$. Over the years, there have been a few more investigations
into this rather old problem \citep{Klafter1996,Berzhkovskii1998,Seki2000,Berezhkovskii2016}.  Of particular interest is the paper by Bicout and Szabo \cite{Bicout1998} who obtained general results for the rate in the case where the opening undergoes non-Markovian fluctuations.  
Even though most of these papers were published long ago, we have
not been able to find in the literature, a simple analytically solvable 
model that accounts for all the experimental results. It is the aim
of this paper to provide such a simple model, based on information from the very interesting
experiments \citep{Yang2003,Kou2004,Min2005,Min2006a}, that have become
available since these original investigations.

Most of the above investigations assume the radius of the opening
to undergo diffusive motion in a harmonic potential. That is, it is an
Ornstein-Uhlenbeck process, with an exponential correlation function.
The major exception to this is the work of Wang and Wolynes \citep{Wolynes1993},
which models the correlation as a stretched exponential, as well as that of Bicout and Szabo \cite{Bicout1998}. In an elegant
set of papers, the group of Xie \citep{Yang2003,Kou2004,Min2005,Min2006a}
investigated the dynamics of the distance between two units in a protein,
that are not directly bonded. Using single molecule fluorescence as
a probe of the distance $x$, they showed that $x$ undergoes subdiffusion,
in which its mean square displacement is proportional to $(time)^{\beta}$
with $\beta<1$. Further, they also showed that the units may be modelled
as being held together by a harmonic spring, and that their motion
is well described by the equation for a subdiffusing Brownian oscillator (see Eq. (\ref{eq:Eqnforsubdiffusion})),
given in Section \ref{sec:subdiffusion}.

Interestingly, the problem of a ``quadratic sink representing a gate
whose dynamics is diffusive'' is of interest in other areas of chemical
physics too. One example is the recent observation of ``anomalous yet Brownian'' diffusion in crowded rearranging media, where the probability distribution of the displacement at short times is found to be exponential, rather than the expected Gaussian \cite{Wang2009}.  It  crosses over to being Gaussian at long times. Interestingly, the mean square displacement at all times is proportional to the time.   This has been explained as resulting from the rearrangement of the medium, leading to  time dependent random changes in the diffusion coefficient of the particle \cite{Chubynsky2014,Metzler2017}. We have suggested a model for calculating the probability
distribution of the position of such a particle.   In this model, the probability
distribution is the Fourier transform of the survival probability
of a particle undergoing Brownian motion \citep{Jain2016,Jain2016a,Jain2017,Jain2017a,Jain2017b}. 
Another interesting study is the ``Fluctating Bottleneck model''
for the passage of a DNA molecule through the $\alpha$-hemolysin pore by Bian
\emph{et al.} \citep{Bian2015,Chatterjee2010}, who found an approximate
solution to the model, using the Wilhemski-Fixman approach. It is
of interest to note that our study provides an exact solution to this
model, though in this paper, we do not discuss our model in this context.

\label{sec:subdiffusion}

\section{The fluctuating Gate Model }

\begin{figure}
\includegraphics[scale=0.4]{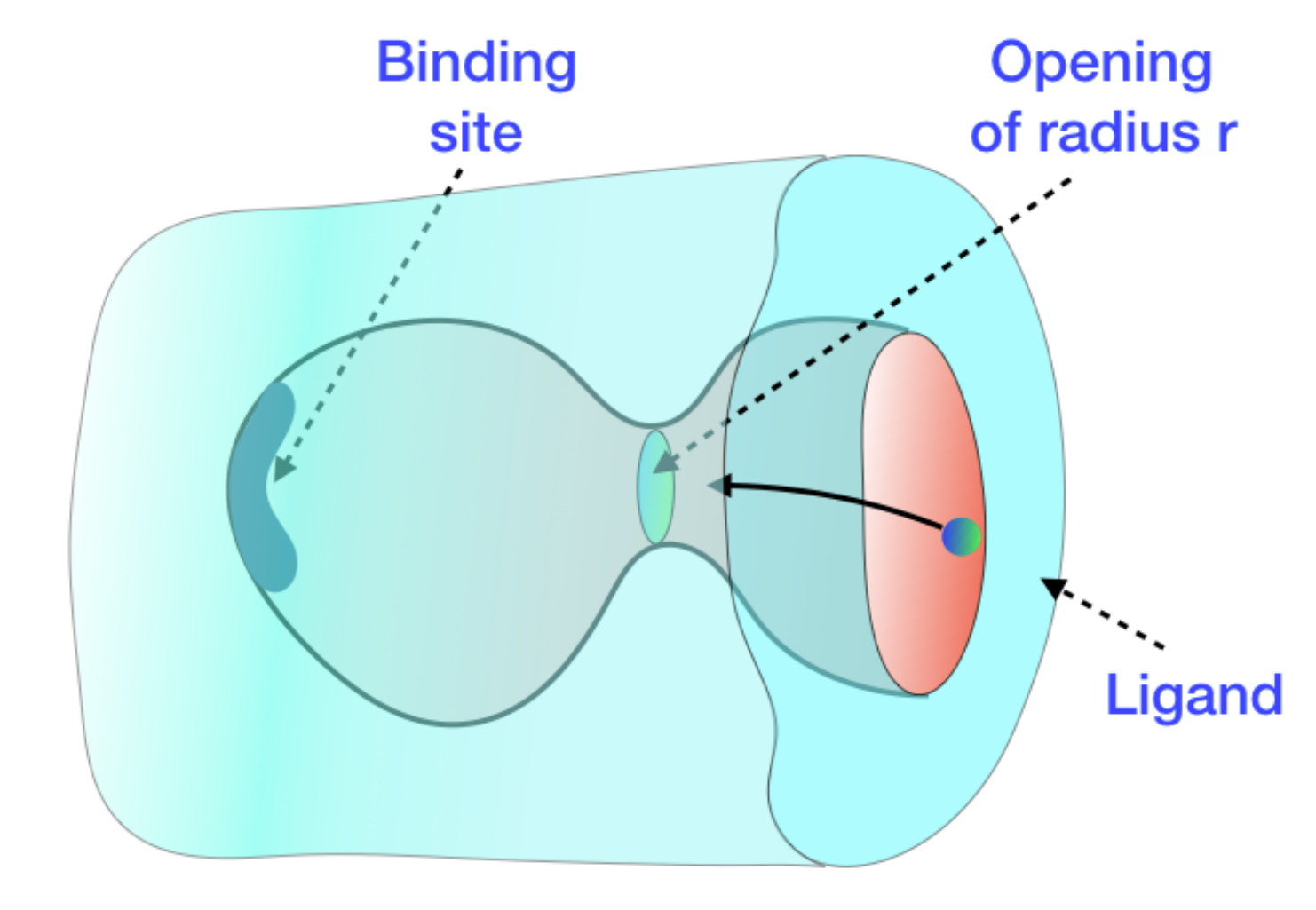}

\caption{Schematic picture of gating}
\label{Opening} 
\end{figure}

The process that we study is shown schematically in Fig. \ref{Opening}.
In order to bind to the site that is inside the cavity, the ligand
has to pass through a gate which is modelled as a circular opening
of radius $r$. The survival probability of the ligand at the time
$t$, $S(t)$ is assumed to obey the equation 
\begin{equation}
\frac{dS(t)}{dt}=-K(r)S(t).\label{eq:Zwanzig-equation}
\end{equation}
Zwanzig \citep{Zwanzig1992} takes the rate of passage of the ligand
to be proportional to $\pi r^{2}$, the area of the circular gate
so that 
\begin{equation}
K(r)=kr^{2}.\label{eq:Rate-dependence-on-r}
\end{equation}
The radius of the pore undergoes thermal fluctuations, obeying the
equation for overdamped motion 
\begin{equation}
\zeta\frac{dr}{dt}=-m\omega_{0}^{2}r+\sqrt{2\zeta k_{B}T}\xi(t),\label{eq:OU-model-Zwanzig}
\end{equation}
where $m$ is the ``mass'' associated with the fluctating co-ordinate,
and $\frac{\zeta}{m\omega_{0}^{2}}$ is its time of relaxation. $\xi(t)$
is Gaussian white noise with mean zero with $\left\langle \xi(t)\xi(t')\right\rangle =\delta(t-t')$
and $T$ is the temperature. $\zeta$ is the friction coefficient
and is directly proportional to the viscosity $\eta$ of the medium.
(Note that our notation is different from that of Zwanzig. Zwanzig's
$\theta$ and $\lambda$ are related to our constants by $\lambda=m\omega_{0}^{2}/\zeta$
and $\theta=\frac{k_{B}T}{m\omega_{0}^{2}}).$ The quantity of interest
is $\left\langle S(t)\right\rangle =\int_{0}^{\infty}dr\bar{S}(r,t)$,
where the noise averaged concentration $\bar{S}(r,t)$ obeys the reaction-diffusion
equation with a sink term that is quadratic in the co-ordinate $r$:
\begin{equation}
\frac{\partial\bar{S}}{\partial t}=-kr^{2}\bar{S}+\frac{k_{B}T}{\zeta}\frac{\partial}{\partial r}\left(\frac{\partial\bar{S}}{\partial r}+\frac{m\omega_{0}^{2}}{k_{B}T}r\bar{S}\right).\label{eq:Zwanzig'sDiffusionEq}
\end{equation}
Assuming that $\bar{S}$ has the equilibrium distribution at the initial
time $t=0$, the instant at which the sink term in Eq. (\ref{eq:Zwanzig'sDiffusionEq})
is switched on, this can be solved exactly. The solution leads to
the following results: (1) The decay of $\left\langle S(t)\right\rangle $
is multi-exponential. (2) At short times, $\bar{S}(t)=1-k\frac{k_{B}T}{m\omega_{0}^{2}}t+O(t^{2})$.
(3) For long times, the decay is single exponential with a rate constant
equal to $k_{eff}=\frac{1}{2}\left(\sqrt{4\frac{k_{B}T}{\zeta}k+(\frac{m\omega_{0}^{2}}{\zeta})^{2}}-\frac{m\omega_{0}^{2}}{\zeta}\right).$
For large $\frac{m\omega_{0}^{2}}{\zeta}$, one gets $k_{eff}\approx(k\frac{k_{B}T}{\zeta})^{1/2}$.
As $\zeta$ is proportional to $\eta,$ we get $k_{eff}\propto\eta^{-1/2}$,
in agreement with experimental observations of Beece \emph{et. al.}
\citep{Beece-1980}, of an $\eta^{-\nu}$ dependence with $\nu$ in
the range $0.4-0.8.$ The deviation of the predicted value of $\nu$
from the observed, shows that the model of Zwanzig can be considered
only as a starting point. This prompted Wang and Wolynes (WW) \citep{Wolynes1993,Wang1994}
to consider a model in which correlation $\left\langle r(t)r(t')\right\rangle $
has the form of a stretched exponential given by $\theta\exp\left[-\left(\lambda|t-t'|\right)^{\beta}\right]$,
with $\beta\leq1$. Note that $\beta=1$ would be the case considerd
by Zwanzig \citep{Zwanzig1992}, and $\beta\leq 1$ is a more general
case of fluctuations, observed in biomolecules and glasses. { In this
more general case, Wang and Wolynes \citep{Wolynes1993} found an
exact expression for the rate constant $k_{eff}$, as a combination of  Hypergeometric functions.  For $\beta=1$, their result reduces to that of Zwanzig.    For $\beta \neq 1$, they obtained an approximation to their result, which predicted  a rate proportional
to $\eta^{-\beta/(1+\beta)}.$} As $\beta\leq1,$ this approximate result gives
$\nu\leq1/2$. They conclude that the inclusion of direct coupling
of the reaction co-ordinate with the viscous medium is required to
have better agreement with experiment.  Bicout and Szabo \cite{Bicout1998} considered the case where the bottleneck undergoes non-Markovian fluctuations.  They obtained analytical expressions for the rate, similar to that of Wang and Wolynes.   Their calculations \cite{Bicout1998} showed the ``Kramers turnover'' behaviour as function of $\eta$, but did not give an explanation of the value of $\nu$ that was reported experimentally.  

Thanks to the advances in single molecule experiments, we now have
more information on the fluctuations of the distance between two sub-units
in a protein. Yang \emph{et. al. }\citep{Yang2003} showed that the
distance $x$ between the Flavin mononucleotide (FMN) and the Flavin
adenine dinucleotide (FAD) in the protein flavin reductase, isolated
from \emph{Escheria coli, }undergoes subdiffusive motion. Its dynamics
is well described by the equation for a subdiffusive Brownian oscillator
(SBO), 
\begin{equation}
\zeta\int_{-\infty}^{t}dt'K_{\alpha}(t-t')\dot{x}(t')=-m\omega_{0}^{2}x(t)+\sqrt{2k_{B}T\zeta}\xi^{\alpha}(t).\label{eq:Eqnforsubdiffusion}
\end{equation}
In the above, $\xi^{\alpha}(t)$ is the fractional Gaussian noise (fGn)
having the correlation function 
\begin{equation}
\left\langle \xi^{\alpha}(t)\xi^{\alpha}(s)\right\rangle =K_{\alpha}(t-s),\label{eq:fGn-Correlation}
\end{equation}
where 
\begin{equation}
K_{\alpha}(t-s)=(\alpha+1)\alpha\left|t-s\right|^{\alpha-1}+2(\alpha+1)|t-s|^{\alpha}\delta(t-s),
\end{equation}
with $0\leq\alpha<1$. $\alpha=0$ corresponds to the usual Gaussian
white noise. Note that Kou and Xie \citep{Kou2004} use the parameter
$H$ instead of our $\alpha$ and the two are related by $H=(\alpha+1)/2.$
The use of such an equation is also justified by the analysis of 
Dua and Adhikari \citep{Dua_2011}. These studies suggest strongly
that a very simple model for the passage through a gate is to assume
that the dynamics of the gate is sub-diffusive. We discuss this model
in the next Section.  The model, as will be seen in Section \ref{rate-constant}, predict a value of $\nu\geq 1/2$, in agreement with the experimental observations.  

\section{Subdiffusive Gating}

\subsection{The Model}

The instantaneous state of the opening may be described by the position
vector $\boldsymbol{r}=x\hat{\boldsymbol{i}}+y\hat{\boldsymbol{j}}$, of a point on its circumference,
with the origin of the co-ordinates located at the center of the pore. The area
of the opening is $\pi r^{2}=\pi \boldsymbol{r}^{2}=\pi\left(x^{2}+y^{2}\right)$. We take $\boldsymbol{r}$
to obey the  subdiffusion equation 
\begin{equation}
\zeta\int_{-\infty}^{t}dt'K_{\alpha}(t-t')\frac{d\boldsymbol{r}(t')}{dt'}=-m\omega_{0}^{2}\boldsymbol{r}(t)+\sqrt{2k_{B}T\zeta}\boldsymbol{\xi}^{\alpha}(t). \label{sub-diff-eqn}
\end{equation}
$\boldsymbol{\xi}^{\alpha}(t)=\xi^{\alpha}_{x}(t)\hat{\boldsymbol{i}}+\xi^{\alpha}_{y}(t)\hat{\boldsymbol{j}}$, where $\xi^{\alpha}_{i}(t)$ with $i=x,y$, are both white noises having the correlation functions $\left <\xi_{i}^{\alpha}(t)\xi_{j}^{\alpha}(s)\right >=\delta_{ij}K_{\alpha}(t-s)$. For more details on this  equation and its application to single molecule
experiments, we refer the reader to the articles by Kou \citep{Kou2008,Kou-AnnRev}.
One can easily calculate the correlation function,
\begin{equation}
C_{\alpha}(t,t')=\left\langle x(t)x(t')\right\rangle =\left\langle y(t)y(t')\right\rangle =\frac{k_{B}T}{m\omega_{0}^{2}}E_{1-\alpha}\left(-(|t-t'|/\tau)^{1-\alpha}\right)\label{eq:Correlation-for-x}
\end{equation}
with 
\begin{equation}
\tau=\left(\frac{\zeta\Gamma(\alpha+2)}{m\omega_{0}^{2}}\right)^{1/(1-\alpha)}\label{eq:tau-zeta-equation}
\end{equation}
and $E_{a}(z)$ is the Mittag-Leffler function, defined by $E_{a}(z)=\sum_{k=0}^{\infty}\frac{z^{k}}{\Gamma(ak+1)}$. Also,
\begin{equation}
\left\langle x(t)y(t')\right\rangle =0. \label{cross-correlation}
\end{equation}
 It is to be noted
that the processes $x(t)$ and $y(t)$ are stationary. Hence we
have $C_{\alpha}(t,t')=C_{\alpha}(t-t').$ As a result of all the above, 
\begin{equation}
\langle \boldsymbol{r}(t)\cdot \boldsymbol{r}(t')\rangle =2 C_{\alpha}(t-t').
\end{equation}
In the following, we will
also need the Fourier cosine transform of $C_{\alpha}(t)$, defined
by 
\begin{eqnarray}
\tilde{C_{\alpha}}(\omega) & =\int_{0}^{\infty}dt\cos\left(t\omega\right)C_{\alpha}(t)=\int_{0}^{\infty}dt\cos\left(t\omega\right)\frac{k_{B}T}{m\omega_{0}^{2}}E_{1-\alpha}\left(-(|t-t'|/\tau)^{1-\alpha}\right)\\
 & =\frac{k_{B}T}{m\omega_{0}^{2}}\;\frac{\tau\cos\left(\frac{\pi\alpha}{2}\right)(\tau w)^{-\alpha}}{1+(\tau w)^{2-2\alpha}+2\sin\left(\frac{\pi\alpha}{2}\right)(\tau w)^{1-\alpha}}.
\end{eqnarray}
See the review by Kou \cite{Kou2008} for the derivation of the correlation function.

\subsection{Survival probability }

On solving Eq. (\ref{eq:Zwanzig-equation}) using Eq. (\ref{eq:Rate-dependence-on-r}),
we get the survival probability of the ligand after a time $t$ to
be 
\begin{equation}
\left\langle S(t)\right\rangle =\left\langle \exp\left(-k\int_{0}^{t}ds\boldsymbol{r}^{2}(s)\right)\right\rangle ,\label{eq:<C(t)>}
\end{equation}
where the average $\left\langle ...\right\rangle $ is over all possible
realizations of $\boldsymbol{r}(s)$. We now introduce $\boldsymbol{\eta}(t)=\eta_{x}(t)\hat{\boldsymbol{i}}+\eta_{y}(t)\hat{\boldsymbol{j}},$
where $\eta_{i}(t)$ with $i=x,y$ are both Gaussian white noises,
having mean zero and correlation $\left\langle \eta_{i}(t)\eta_{j}(t')\right\rangle =\delta_{ij}\delta(t-t')$.  Using the result, 
\[
\int D\boldsymbol{\eta}(s) \exp\left(-\int_{0}^{t}ds\boldsymbol{\eta}^{2}(s)+2i\sqrt{k}\int_{0}^{t}ds\boldsymbol{r}(s)\cdot\boldsymbol{\eta}(s)\right)=  \exp\left(-k\int_{0}^{t}ds\boldsymbol{r}^{2}(s)\right),
\]
it is possible to rewrite Eq. (\ref{eq:<C(t)>}) as 
\[
\left\langle S(t)\right\rangle =\int D\boldsymbol{\eta}(s)\left\langle \exp\left(-\int_{0}^{t}ds\boldsymbol{\eta}^{2}(s)+2i\sqrt{k}\int_{0}^{t}ds\boldsymbol{r}(s)\cdot\boldsymbol{\eta}(s)\right)\right\rangle ,
\]
where the functional integral $\int D\boldsymbol{\eta}(s)$ is over
all possible realizations of $\boldsymbol{\eta}(s)$. As $\boldsymbol{r}(s)$
is a Gaussian stochastic process with mean zero, the average over
it is easily performed to get (see the book by Chaichian and Demichev \cite{Chaichian-Demichev} or the book by Zinn-Justin \cite{Zinn-Justin2005}, for an introduction to functional integrals and their applications). 
\begin{eqnarray}
\left\langle S(t)\right\rangle  & = & \int D\boldsymbol{\eta}(s)\exp\left(-\int_{0}^{t}ds\boldsymbol{\eta}^{2}(s)-2k\int_{0}^{t}ds'\int_{0}^{t}ds\boldsymbol{\eta}(s)\cdot\left\langle \boldsymbol{r}(s)\;\boldsymbol{r}(s')\right\rangle \cdot \boldsymbol{\eta}(s')\right) 
\end{eqnarray}
The tensorial average $\left\langle \boldsymbol{r}(s)\;\boldsymbol{r}(s')\right\rangle $ is to be performed over all realisations of the process $\boldsymbol{r}(s)$.  Using Eq. (\ref{eq:Correlation-for-x}) and (\ref{cross-correlation}), this may be written as 
\begin{eqnarray}
 \left\langle S(t)\right\rangle & = & \int D\boldsymbol{\eta}(s)\exp\left(-\int_{0}^{t}ds\boldsymbol{\eta}^{2}(s)-2k\int_{0}^{t}ds'\int_{0}^{t}ds\boldsymbol{\eta}(s)\cdot \boldsymbol{\eta}(s')C_{\alpha}(s-s')\right).\end{eqnarray}
 Noting that $\boldsymbol{\eta}(s)$ is a two dimensional vector, and  performing the functional integration over it gives  
 \begin{eqnarray}
 \left\langle S(t)\right\rangle & = & A\left(\det\left[\delta(s-s')+2kC_{\alpha}(s-s')\right]\right)^{-1}\hspace{0.5cm}\mbox{with}\hspace{0.5cm}s,s'\;\in\;(0,t).\label{eq:Survival-probability0}
\end{eqnarray}
In the above, $\delta(s-s')+2kC_{\alpha}(s-s')$ is the $(s,s')^{th}$ element of a functional matrix
whose labels $s,s'$ are continuous. $A$ is a constant, equal to $\det\left[\delta(s-s')\right],$
which is divergent. Note that the value of $\left\langle S(t)\right\rangle $
is the ratio of two divergent quantities and is always finite, as
it should be (see below). We use the identity $\det B=\exp\left(Tr\ln B\right),$
where $Tr$ stands for the trace of the matrix, valid for any Hermitian
matrix $B,$ to write

\[
\left\langle S(t)\right\rangle =A\exp\left(-tr\ln\left[\delta(s-s')+2kC_{\alpha}(s-s')\right]\right).
\]
Expanding the logaritham, and using the condition $\left\langle S(t)\right\rangle |_{k=0}=1,$
we get 
\begin{equation}
\left\langle S(t)\right\rangle =\exp\left(\sum_{n=1}^{\infty}(-1)^{n}\frac{(2k)^{n}}{n}Tr\boldsymbol{C}_{\alpha}^{n}\right),\label{eq:Survival-probability1}
\end{equation}
where $\boldsymbol{C}_{\alpha}$ is the matrix with continuous labels
$s,s'$, defined by $\boldsymbol{C}_{ \alpha}(s,s')=C_{\alpha}(s-s').$
Equation (\ref{eq:Survival-probability1}) is our final expression,
and we can now analyse it, to get the detailed behavior of the survival
probability.

\subsection{Short time behavior}

For small times, i.e., $t\ll\tau,$ we can approximate $C_{\alpha}(s-s')$
by $C_{\alpha}(0).$ Doing this in each term in Eq. (\ref{eq:Survival-probability1}),
noting from Eq. (\ref{eq:Correlation-for-x}) that $C_{\alpha}(0)=\frac{k_{B}T}{m\omega_{0}^{2}}$,
and summing the resultant series gives 
\[
\left\langle S(t)\right\rangle =\left(1+2k\frac{k_{B}T}{m\omega_{0}^{2}}t\right)^{-1}\approx1-2k\frac{k_{B}T}{m\omega_{0}^{2}}t+O(t^{2}),
\]
exactly as in the case of the Zwanzig model. Thus subdiffusion of the
gate does not make any difference to the short term behavior of the
survival probability.

\subsection{An exact expression for numerical evaluation of
$\left\langle S(t)\right\rangle $} 

An exact expression, which may be used for numerical evaluation of
the survival probabilty can be obtained by discretisation of the time
interval $(0,t)$ into $N$ discrete intervals each of duration $\Delta t$,
so that $N\Delta t=t.$ Denoting $s_{j}=(j-1)\Delta t,$ with $j=1,2,3,\ldots N$,
and approximating the matrix $\boldsymbol{C}_{\alpha}(s,s')$ by
the finite dimensional matrix $\boldsymbol{C}_{\alpha;N}$ with
matrix elements, $\left(C_{\alpha;N}\right)_{ij}=C_{\alpha}(s_{i}-s_{j})\Delta t$
in each term of the sum in the exponent of Eq. (\ref{eq:Survival-probability1})
gives        
\[
\left\langle S(t)\right\rangle =\lim_{N\rightarrow\infty,\Delta t\rightarrow0,N\Delta t=t}\;\left(\det[\boldsymbol{I+}2k\boldsymbol{C_{\alpha}}_{N}]\right)^{-1},
\]
where $\boldsymbol{I}$ is the $N\times N$ identity matrix. The value
of $N$ can be chosen sufficiently large to get the survival probability
to any desired accuracy.

\subsection{Survival probability in the long time limit}

One can easily get an approximation for $\left\langle S(t)\right\rangle $
in the long time limit. First we note that the matrix $\boldsymbol{I+}\boldsymbol{C}_{\alpha;N}$
would be the Hamiltonian matrix for a chain of $N$ atoms, in a tight binding model, having
one orbital on each atom, with all the diagonal elements equal to $1+2kC_{\alpha}(0)\Delta t$
and the $ij^{th}$ offdiagonal element equal to $2k C_{\alpha}(\Delta t(i-j))\Delta t.$
We also note that $C_{\alpha}(t)$ is a decaying function of $t,$
decreasing like $t^{\alpha-1}$ for large values of $t,$ for $\alpha\neq0$
(if $\alpha=0,$ then it decays exponentially). In either case, for
large values of $t$ one expects that modifying the Hamiltonian matrix
by imposing periodic boundary conditions on the chain of atoms will
not cause a significant change to its eigenvalues. Once the condition
is imposed, the eigenvectors of the matrix are $\phi_{j}$  with its $n^{th}$ element being given by $\phi_{j,n}=\frac{1}{\sqrt{N}}e^{i\frac{2\pi n j}{N}}$, where $j$ varies from $-N/2$ to $N/2+1$ and $n$ varies from $0$ to $(N-1)$.  With this, it is easy to calculate the $j^{th}$ eigenvalue of the matrix $\boldsymbol{I+}\boldsymbol{C}_{\alpha;N}$.  It is given by
\begin{equation}
\epsilon_{j}=1+2k\Delta t\sum_{n=-N/2+1}^{N/2}C_{\alpha}(n\Delta t)\,\exp\left(i\frac{2\pi nj}{N}\right)\hspace{1em}\mbox{ with }j=0,\pm1,\pm2,...\label{eq:jth eigenvalue}
\end{equation}
In the limit $\Delta t\rightarrow0$, and $N\rightarrow\infty$, with
$N\Delta t=t$,  one can convert the sum over $n$ to integration.  This gives 
\[
\epsilon_{j}=1+2k\int_{-t/2}^{t/2}ds\,C_{\alpha}(s)\,\exp\left(i\frac{2\pi js}{t}\right).
\]
Noting that $C_{\alpha}(s)$ is an even function of $s$, the above may be written as   
\[
\epsilon_{j}=1+2k\int_{-t/2}^{t/2}ds\,C_{\alpha}(s)\,\cos\left(\frac{2\pi js}{t}\right).
\]
This may be re-written as
\begin{equation}
\epsilon_{j}\simeq1+4k\int_{0}^{t/2}ds\,\cos\left(\frac{2\pi js}{t}\right)\,C_{\alpha}(s). \label{jth-eigenvalue}
\end{equation}
For large times, the upper limit of integration can be replaced
with infinity, and then 
\begin{equation}
\epsilon_{j}\simeq1+4k\tilde{C_{\alpha}}(\frac{2\pi j}{t})\hspace{1em}\mbox{for \ensuremath{j\neq0}}. \label{epsilon-j}
\end{equation}
The case where $j=0$ needs special attention, and is evaluated below:

\begin{equation}
\epsilon_{0}=1+4k\frac{k_{B}T}{m\omega_{0}^{2}}\int_{0}^{t/2}dsE_{1-\alpha}\left(-(s/\tau)^{1-\alpha}\right).\label{zeroeigenvalue}
\end{equation}
For $t\rightarrow\infty,$ the major contribution to the integral
comes from large values of $s$ at which one may use the asymptotic
expression $E_{1-\alpha}(-(s/\tau)^{1-\alpha})=\frac{(s/\tau)^{\alpha-1}}{\Gamma(\alpha)}+O((s/\tau)^{2\alpha-2}).$
This gives 
\begin{equation}
\epsilon_{0}\approx1+4k\frac{k_{B}T\tau}{m\omega_{0}^{2}\Gamma(\alpha+1)}\left(\frac{t}{2\tau}\right)^{\alpha}%\approx2k\frac{k_{B}T\tau}{m\omega_{0}^{2}\Gamma(\alpha+1)}\left(\frac{t}{2\tau}\right)^{\alpha}.
\label{eq:epsilon0}
\end{equation}
Taking into account of the fact that $\epsilon_{0}$ is non-degenerate,
and that all other eigenvalues are doubly degenerate, we get 
\begin{equation}
\left\langle S(t)\right\rangle =\left(\det[\boldsymbol{I}+4k\boldsymbol{C_{\alpha}}_{N}]\right)^{-1}=\frac{1}{\epsilon_{0}\left(\prod_{j=1}^{\infty}\epsilon_{j}\right)^{2}},
\end{equation}
which may be approximated as {\color{blue} %(Explain the derivation below)
}
\begin{equation}
\left\langle S(t)\right\rangle  =  \exp\left[-\sum_{j=-\infty}^{\infty}\ln(\epsilon_{j})\right].
\end{equation}
Using the expression for $\epsilon_{j}$ in Eq. (\ref{epsilon-j}) and (\ref{eq:epsilon0}), we get
\begin{equation}
  \left\langle S(t)\right\rangle   =  \exp\left[-\sum_{j=-\infty}^{\infty}\ln\left(1+4k\tilde{C_{\alpha}}(\frac{2\pi|j|}{t})\right)\right].\nonumber \end{equation}
 In the limit of large $t$, it is a good approximation to replace the sum by integration.  Introducing $\omega =2\pi j/t$ and converting the sum over $j$ to integration over $\omega$, we can write
 \begin{equation}
 \left\langle S(t)\right\rangle =  \exp\left[-\frac{t}{\pi\tau}\int_{0}^{\infty}d\omega\ln\left(1+4k\tau\frac{k_{B}T}{m\omega_{0}^{2}}\;\frac{\cos\left(\frac{\pi\alpha}{2}\right)\omega^{-\alpha}}{(1+\omega^{2-2\alpha}+2\sin\left(\frac{\pi\alpha}{2}\right)\omega^{1-\alpha})}\right)\right].\label{integralforsurvival}
\end{equation}
Notice that the integrand is divergent at $\omega=0$, this being a result
of the divergence of the $j=0$ eigenvalue in the $t\rightarrow\infty$
limit. This does not cause any particular problem, as the integral
in Eq. (\ref{integralforsurvival}) is well behaved. 

\begin{figure}[h]
\centering \includegraphics[width=0.7\linewidth]{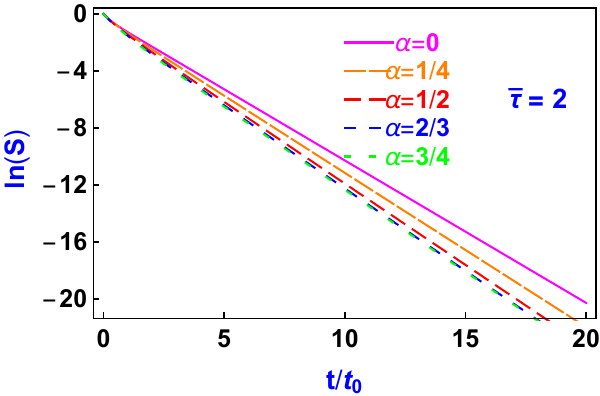} \caption{Plots of survival probability as a function of time $\bar{t}=t/t_{0}$
for different values of $\alpha$. $\bar{\tau}=\tau/t_{0}$ is taken
to be $2$. The plots show that the decay is exponential for long
times. Slopes obtained by fitting the data are given in Table \ref{rates1}.}
\label{fig2:ln-survival} 
\end{figure}

\begin{figure}[h]
\centering \includegraphics[width=0.7\linewidth]{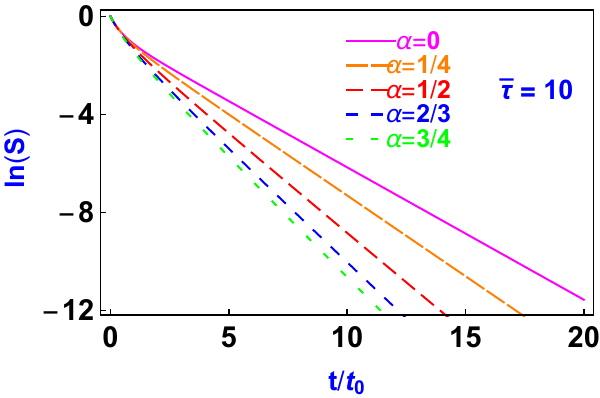} \caption{Plots of survival probability as a function of time $\bar{t}=t/t_{0}$
for different values of $\alpha$. $\bar{\tau}=\tau/t_{0}$ is taken
to be $10$. The plots show that the decay is exponential for long
times. }
\label{fig3:ln-survival} 
\end{figure}

It is convenient
to introduce the dimensionless variables $\bar{t}=t/t_{0}$, where
$t_{0}=\frac{m\omega_{0}^{2}}{k_{B}T}$ and $\bar{\tau}=\frac{\tau}{t_{0}}$.
Then, the survival probability $S$ is a function
of these two reduced variables alone. Plots of this function are 
shown in Figures \ref{fig2:ln-survival} and \ref{fig3:ln-survival}.
We have also compared the values of $k_{eff}$ obtained by fitting
the numerical results with $k_{eff}$ calculated numerically using
equation (\ref{integralforsurvival}) in Tables \ref{rates1} and
\ref{rates2}. It is to be noted that in general the agreement is
good for small values of $\alpha$ $(\leq1/2)$ but not so for higher
values of $\alpha$. See the result for $\alpha=3/4$ in table \ref{rates2}.
For higher values of $\alpha$ the Mittag-Leffler function
$E_{1-\alpha}(-(t/\tau)^{1-\alpha})\sim\frac{(t/\tau)^{\alpha-1}}{\Gamma(\alpha)},$ which
means that $C_{\alpha}(t)$ decays very slowly with and hence our
approximation of using periodic boundary condition becomes poorer as
the value of $\alpha$ approaches unity.

\subsection{The effective rate constant, $k_{eff}$}
\label{sub-1}

\begin{table}[h]
\centering{}%
\begin{tabular}{|c|c|c|c|}
\hline 
$\hspace{0.3cm}\alpha \hspace{0.3cm}$  & $\bar{k}_{eff}$ (fitted)  & $\bar{k}_{eff}$ (from Eq. (\ref{eq:bark_eff})) \\
\hline 
$0$  & $0.9997$  & $0.9999$  \\ 
\hline 
$\frac{1}{4}$  & $1.0821$  & $1.0824$  \\ 
\hline 
$\frac{1}{2}$ & $1.1404$ & $1.1398$ \\ 
\hline 
$\frac{2}{3}$  & $1.1544$  & $1.1398$ \\ 
\hline 
$\frac{3}{4}$  & $1.1491$  & $1.0968$ \\ 
\hline 
\end{tabular}\caption{Values of $k_{eff}$ for different values of $\alpha$. All the results
are for $\bar{\tau}=2$.}
\label{rates1} 
\end{table}
\begin{table}[h]
\begin{centering}
\begin{tabular}{|c|c|c|c|}
\hline 
$\hspace{0.3cm}\alpha$ \hspace{0.3cm} & $\bar{k}_{eff}$ (fitted)  & $\bar{k}_{eff}$ (from Eq. (\ref{eq:bark_eff}))  \tabularnewline
\hline 
$0$  & $0.5402$  & $0.5403$   \\
\hline 
$\frac{1}{4}$ & $0.6586$  & $0.6575$   \\
\hline 
$\frac{1}{2}$ & $0.8052$  & $0.8000$   \\
\hline 
$\frac{2}{3}$  & $0.9127$  & $0.8910$  \\
\hline 
$\frac{3}{4}$  & $0.9634$  & $0.9038$  \\
\hline 
\end{tabular}
\end{centering}
\caption{Values of $\bar{k}_{eff}$ for different values of $\alpha$. All
the results are for $\bar{\tau}=10$. It is to be noted that the agreement
is good for smaller values of $\alpha$, but for $\alpha=3/4$, the
agreement is not so good. } \label{rates2}
\end{table}
\begin{figure}
\includegraphics[width=0.7\linewidth]{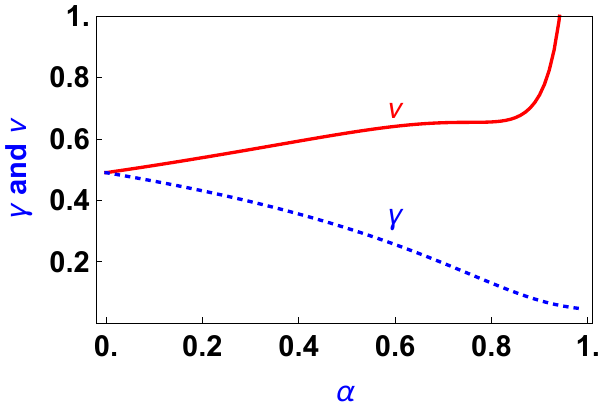}
\caption{Plot of $\gamma$ and $\nu$ against $\alpha$.} \label{Fig4-nuvsalp}
\end{figure}
It is clear that the long term decay is exponential, being of the
form $e^{-k_{eff}t}$. From Eq. (\ref{integralforsurvival}), an approximation to $k_{eff}$ is 
\begin{equation}
k_{eff}=\frac{1}{\pi\tau}\int_{0}^{\infty}d\omega\ln\left(1+4k\tau\frac{k_{B}T}{m\omega_{0}^{2}}\;\frac{\cos\left(\frac{\pi\alpha}{2}\right)\omega^{-\alpha}}{(1+\omega^{2-2\alpha}+2\sin\left(\frac{\pi\alpha}{2}\right)\omega^{1-\alpha})}\right),\label{eq:k_eff}
\end{equation}
which in terms of the dimension-less quantities $\bar{t}$, $\bar{\tau}$
and $\bar{k}_{eff}=k_{eff}t_{0}$ may be written as 
\begin{equation}
\bar{k}_{eff}=\frac{1}{\pi\bar{\tau}}\int_{0}^{\infty}d\omega\ln\left(1+4\bar{\tau}\;\frac{\cos\left(\frac{\pi\alpha}{2}\right)\omega^{-\alpha}}{(1+\omega^{2-2\alpha}+2\sin\left(\frac{\pi\alpha}{2}\right)\omega^{1-\alpha})}\right).\label{eq:bark_eff}
\end{equation}
For $\alpha=0$, the integral can be evaluated exactly, to get the result
\begin{equation}
\bar{k}_{eff}=\frac{1}{\bar{\tau}}\left(\sqrt{1+4\bar{\tau}}-1\right),
\end{equation}
which is the same as that obtained by Zwanzig \citep{Zwanzig1992}
(Note that while Zwanzig considers one dimensional version of the
problem, while our analysis is for a two dimensional opening, and
hence Zwanzig's rate is only half of ours). On taking $\tau\propto\zeta\propto\eta$,
for large values of $\frac{k_{B}T}{m\omega_{0}^{2}}$ (i.e., $\bar{\tau}\gg 1$), we find $k_{eff}\propto\eta^{-1/2},$ which
is the result of Zwanzig. We now consider the situation where the
radius of the opening undergoes subdiffusion. For $\alpha>0,$ we
have not been able to evaluate $k_{eff}$ analytically. Hence we calculate
the integral in Eq. (\ref{eq:k_eff}) numerically, and find its dependence
on $\zeta.$ We took $\bar{\tau}$ to be large and to vary from $1000$
to $10,000$. The value of $\overline{k}_{eff}$ was calculated for
each $\bar{\tau}$, using Eq. (\ref{eq:k_eff}) and MATHEMATICA. For
each value of $\alpha$, plots of $\ln\bar{k}_{eff}$ against $\ln\bar{\tau}$
were then made, and were found to be linear with negative slope. This
implies $\bar{k}_{eff}\propto(\bar{\tau})^{-\gamma}.$ As $\bar{\tau}\propto\zeta^{1/(1-a)}$
(see Eq. (\ref{eq:tau-zeta-equation})), we get $\bar{\tau}\propto\eta^{-\nu}$,
with $\nu=\gamma/(1-\alpha)$. The results for $\gamma$ and $\nu$
are plotted in Fig. \ref{4-nuvsalp}, for various values of $\alpha$.
It is clear that the values range from $0.5$ to greater than unity
(it is greater than unity for values of $\alpha$ very close to $1$).
Remembering that $\zeta$ is proportional to the viscosity $\eta$,
this is in better agreement with the experiments, than the simple
diffusive model to Zwanzig \citep{Zwanzig1992}, or the stretched
exponential model of Wang and Wolynes \citep{Wolynes1993}. 
\label{rate-constant}    
\section{Summary and Conclusions}
We have found exact solution to the problem of the survival probability of a 
Brownian oscillator that undergoes sub-diffusion, moving in presence of a quadratic sink. This is a problem
of great interest in different areas of chemical physics \cite{Zwanzig1992,Bian2015,Wolynes1993}
and for which only approximate solutions were known  \cite{Bian2015}.
Our solution was used to analyse the problem of ligand passage through
a fluctuating bottle neck. It was found that the model predicts an
effective rate constant proportional to $\eta^{-\nu}$, $\eta$ being
the viscosity of the medium, with a $\nu\geq1/2$. This is in better agreement with experiments ($\nu$ in the
range $0.4$ to $0.8$) than the previous models, which predict $\nu\leq1/2$
\cite{Zwanzig1992,Wolynes1993}.  Hence sub-diffusion \cite{Kou2004,Zheng2015a} of the opening can explain the experimental observations \cite{Beece-1980,Zwanzig1992} on the viscosity dependence of ligand binding to a protein.

{ As already pointed out, there is experimental evidence for sub-diffusive motion \citep{Yang2003,Kou2004,Min2005,Min2006a} of the distance between two parts of a protein that are not directly bonded.  There is also theoretical evidence for the same, for a simple polymer model \citep{Dua_2011}.  It would be nice to have more experimental and theoretical evidence in this direction.  An attractive possibility would be to analyse the dynamics of the area of a gate using molecular dynamics simulations, as in \citep{Nury2010}.}

\section{Acknowledgements}

I  thank Prof. Samuel C Kou (Harvard University) for making me aware of, and sending
me a copy of the reference \citep{Kou-AnnRev}. This work was supported
by the Department of Science and Technology, Government of India,
through the J.C. Bose Fellowship.  

\bibliographystyle{apsrev}

%\bibliography{subdiffusiveGate}
\end{document}